\documentclass[9pt,twocolumn,twoside]{pnas-new}
\templatetype{pnasresearcharticle} 
\usepackage{color}
\definecolor{LightGreen}{cmyk}{0.68, 0.0, 0.73, 0.37}
\definecolor{Black}{cmyk}{1.0, 1.0, 1.0, 1.0}
\newcommand{\change}[1]{\color{Black}{#1}}

\newcounter{MD}
\newcommand*\MD%
{%
\ifnum\theMD>0%
{MD}\fi%
\ifnum\theMD=0%
{Molecular Dynamics (MD) \setcounter{MD}{1}}\fi%
}

\newcounter{SI}
\newcommand*\SI%
{%
\ifnum\theSI>0%
{SI}\fi%
\ifnum\theSI=0%
{Supplementary Information (SI) \setcounter{SI}{1}}\fi%
}

\begin{document}
\nolinenumbers

\title{Programmable patchy particles for materials design}

\author[a,1]{Ella M. King}
\author[b,c,1]{Chrisy Xiyu Du}
\author[b]{Qian-Ze Zhu}
\author[d,e]{Samuel S. Schoenholz}
\author[b,d]{Michael P. Brenner}

\affil[a]{Department of Physics, Harvard University, Cambridge MA 02139, USA}
\affil[b]{School of Engineering and Applied Sciences, Harvard University, Cambridge MA 02139, USA}
\affil[c]{Mechanical Engineering, University of Hawai`i at Mānoa, Honolulu HI 96822, USA}
\affil[d]{Google Research, Mountainview CA 94043}
\affil[e]{OpenAI}

\leadauthor{King}
\leadauthor{Du}

\significancestatement{The development of new materials has been a transformative force in shaping the modern world. The traditional approach to creating new functional materials relies on a combination of hard-won intuition and arduous labor to sift through the innumerable possibilities. However, as the functions we need grow increasingly complex, finding that rare substance amid all possible materials becomes increasingly difficult. Instead, we will need methods for the direct design of novel materials. Existing methods for inverse materials design are limited to extremely simple components. Here, we dramatically increase the complexity of the designable components, broadening the scope of materials we can design.}

\authorcontributions{Please provide details of author contributions here.}
\authordeclaration{The authors declare no conflict of interest here.}
\equalauthors{\textsuperscript{1}Ella M. King contributed equally to this work with Chrisy Xiyu Du}
\correspondingauthor{\textsuperscript{2}To whom correspondence should be addressed. E-mail: mpbrenner@seas.harvard.edu}


\begin{abstract}
Direct design of complex functional materials would revolutionize technologies ranging from printable organs to novel clean energy devices. However, even incremental steps towards designing functional materials have proven challenging. If the material is constructed from highly complex components, the design space of materials properties rapidly becomes too {\change{computationally expensive}} to search. On the other hand, very simple components such as uniform spherical particles are not powerful enough to capture rich functional behavior. Here, we introduce a differentiable materials design model with components that are simple enough to design yet powerful enough to capture complex materials properties: rigid bodies composed of spherical particles with directional interactions (patchy particles). We showcase the method with self-assembly designs ranging from open lattices to self-limiting clusters, all of which are notoriously challenging design goals to achieve using purely isotropic particles. By directly optimizing over the location and interaction of the patches on patchy particles using gradient descent, we dramatically reduce the computation time for finding the optimal building blocks. 
\end{abstract}

\dates{This manuscript was compiled on \today}
\doi{\url{www.pnas.org/cgi/doi/10.1073/pnas.XXXXXXXXXX}}

\maketitle
\thispagestyle{firststyle}
\ifthenelse{\boolean{shortarticle}}{\ifthenelse{\boolean{singlecolumn}}{\abscontentformatted}{\abscontent}}{}

\nolinenumbers

\dropcap{S}ignificant efforts have been made towards designing synthetic materials that rival the complexity we observe in biological systems \cite{sindoro2014colloidal, jones2015programmable, rogers2016using, zeravcic2017colloquium, hagan2021equilibrium, hueckel2021total, du2022programming}. However, many of the synthetic systems studied suffer from one of two fatal flaws: either the system is too simple to be able to replicate complex behaviors, or the system is too complex to be easily designable. {\change{By combining the principles that enable machine learning methods to  efficiently navigate large parameter spaces with physics- and materials science-informed models, we introduce a system that both complex enough to capture desirable functional behavior and is amenable to inverse design.}}

One major line of inquiry towards designing complex functional materials focuses on materials with uniform spherical particles as components. {\change{While this approach has led to promising advances \cite{dshemuchadse2021moving, fang2020two, pineros2018inverse, lindquist2018inverse, adorf2018inverse}}}, the absence of directional interactions significantly limits the design space. In materials with spherical components, designed interactions either are too complicated to be experimentally realizable \cite{dshemuchadse2021moving}, or necessitate that every particle interact uniquely with every other particle in the system, which cannot be physically instantiated at scale \cite{angioletti2016theory}. Conversely, a rich literature of work \cite{ouldridge2011structural, bowers2006scalable, angioletti2016theory, wang2004development} has been conducted by running forward simulations of systems with highly complex components, such as proteins. {\change{The complexity of the components means individual simulations are extremely computationally intensive, making inverse design approaches untenable. Moreover, the design space for these systems is too vast to search effectively.}}

Breaking rotational symmetry of the component particles vastly increases the potential for materials designability without relying on having a large number of particle types. Extensive research has been done on anisotropic particles, ranging from mapping out phase diagrams for hard particles with non-trivial shapes \cite{klotsa2018intermediate} to designing patchy particles to self-assemble {\change{open lattices such as}} cubic diamond structures \cite{romano2020designing, rao2020leveraging}. The design space for anisotropic particles is immense \cite{glotzer2007anisotropy}, making brute force approaches to searching for desirable properties unsustainable for modern materials design.  {\change{However, the dramatic advances brought about by the machine learning community have recently made it possible to search vast design landscapes for regions with desirable behavior.}} 

{\change{We introduce the first inverse design method for anisotropic particles that is both physics-based and fully differentiable. Our work has parallels to previous work in that we design interactions that lead to target self-assembled structures~\cite{fang2020two, pineros2018inverse, lindquist2018inverse, adorf2018inverse, whitelam2021neuroevolutionary, dshemuchadse2021moving}. However, many of these approaches lead to highly complex interactions ~\cite{whitelam2021neuroevolutionary, dshemuchadse2021moving, angioletti2016theory} which are inaccessible experimentally. In contrast, we move the complexity from the interactions to the component particle geometries, which may lead to more experimentally realizable designs. There have been encouraging initial efforts to inverse-design materials with anisotropic components, but existing methods are either highly system-dependent \cite{geng2019engineering, zhang2017optimal} require large training data sets for each system of interest \cite{long2018rational, roding2022inverse}, or necessitate that the complex system can be captured by a much simpler, lower dimensional representation \cite{ma2019inverse, pinto2023design}. Our method is system independent, does not rely on large amounts of training data, and can capture complex systems that do not exhibit lower dimensional representations.}}

{\change{Here, we introduce a framework that capitalizes on the advances wrought by machine learning to broadly enable inverse design of complex, anisotropic systems.}} 
This framework {\change{end-to end differentiable}} and is system-independent within JAX-MD \cite{schoenholz2020jax}, a \MD\ engine with automatic differentiation (AD) \cite{baydin2018automatic} enabled. AD is the workhorse underlying the explosion in productivity in the machine learning community in recent decades. 
We introduce the ability to directly optimize over particle geometry and anisotropic interactions. We demonstrate the method specifically on patchy particles, a model system that is simple enough to design, yet rich enough to capture features such as directional interactions.  In this paper, we first discuss the implementation of the method, and then we demonstrate the versatility of the platform by showcasing three examples: (i) stabilization of a Kagome lattice, (ii) self-limiting 2D ring assembly and (iii) stabilization of 3D finite clusters. {\change We include a python notebook \cite{notebook_git} that demonstrates how to perform the optimization for the 3D finite cluster case.} Direct gradient descent of building block properties will enable researchers to efficiently design novel  materials with targeted properties and functions.

\begin{figure}
    \centering
    \includegraphics[width = 1.0\linewidth]{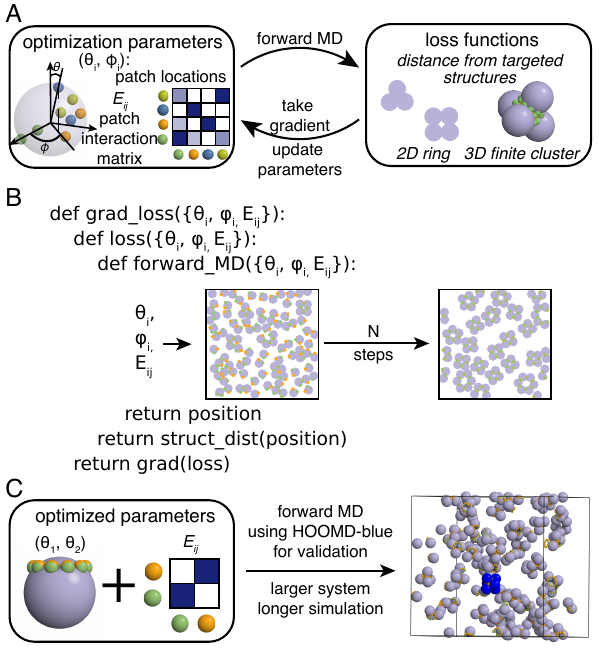}
    \caption{A. \textbf{Optimizing patchy particle interactions} The optimization parameters (shown on the top) are the patch locations and the interaction matrix of patch strengths. These parameters are used to run a forward simulation, and a loss function is computed. We then take the gradient of the loss function with respect to the optimization parameters, and update the parameters accordingly. The loss functions vary for different optimization targets. B. \textbf{Gradient of Loss Function respect to parameters for optimization} The pseudo code demonstrates how the gradient is computed based on the parameters for optimization.
    C. \textbf{Extrapolation to more performant MD engines} We test optimal parameters in HOOMD-blue, showing both that optimal parameters are valid across different MD engines and enabling rapid testing for longer simulations with more particles.}
    \label{fig:model}
\end{figure}

\section*{Method \label{sec:Method}}
\MD\ simulations are a powerful tool for understanding micro- and nanoscale systems. When combined with inverse design methods, \MD\ simulations can be used to design highly complex structures and materials properties, ranging from intricate crystal structures \cite{van2015digital, geng2019engineering, lindquist2018inverse}, finite clusters \cite{long2018rational}, phase transitions \cite{du2020inverse}, and kinetics \cite{goodrich2021designing} in self-assembled systems. 

Existing methods that combine \MD\ with automatic differentiation are limited to simulations of isotropic particles \cite{goodrich2021designing}, which significantly limits the design space of complex materials functions. However, many standard \MD\ libraries, such as HOOMD-blue \cite{anderson2020hoomd} and LAMMPS \cite{LAMMPS}, offer support for simulations of non-isotropic objects called rigid bodies. Rigid bodies in \MD\ are generally defined as a system of spheres with no internal degrees of freedom, which can be used to simulate building blocks that have arbitrary shape and directional interactions.

Here, we build on an existing software package, JAX-MD~\cite{schoenholz2020jax}, that enables fully differentiable \MD\ simulations. The original release of JAX-MD did not support building blocks or integrators with rotational degrees of freedom. Here, we enable both simulation and differentiation of systems with anisotropic particles. 

To simulate and optimize over anisotropic particles, we extended all the available integrators in JAX-MD to account for rotational degrees of freedom, following the algorithm introduced in \cite{miller2002symplectic}. The technical details of threading the gradients through simulations of anisotropic particles, in addition to a detailed discussion of other features of the implementation, can be found in the \SI.

Our prototypical optimization procedure for inverse design with \MD\ begins with specifying all the variables needed for a standard forward simulation. These variables include the system information, such as the number of particles, pair potential, and box size, as well as the integrator information, such as the integrator type, temperature, and step size. While any of these variables can be optimized, we focus on examples where we optimize over only the pair potentials and the particle geometries. We parameterize the pair potential by a matrix of interaction strengths, and we parameterize the particle geometries by the locations of patches on central particles (Fig.~\ref{fig:model}A).

To compute gradients of patch locations $(\theta_i, \phi_i)$ and interaction strengths ($E_{ij}$), we first define a forward \MD\ simulation function (Fig.~\ref{fig:model}B) that takes $(\theta_i, \phi_i, E_{ij})$ as input parameters and returns the position data of every particle in the simulation.  We then use the position data as input for our loss function, where we define criteria to evaluate whether the system is close to our targeted behavior.  Lastly, we take the gradient of the loss function respect to $(\theta_i, \phi_i, E_{ij})$, and update $(\theta_i, \phi_i, E_{ij})$ based on the gradient values for the next iteration.  We use the Adam optimizer to update our parameters based on the gradient computation. 

As a final step, shown in Fig.~\ref{fig:model}C, we use our optimal parameters to run forward simulations in a more performant \MD\ engine. These forward simulations are run for longer timescales and with more particles, demonstrating validity of our optimal parameters across different \MD\ engines and beyond the time- and length-scale of the simulations we optimize over. This iterative procedure shows one possible means of bringing inverse-design into traditional \MD\ simulation.

We note that this optimization procedure does not rely on any black-box methods: all the functions we differentiate are physics-based simulations. Because every step of the optimization function, from \MD\ simulation to loss function evaluation, is fully differentiable, we are able to compute gradients of arbitrary parameters.

\begin{figure}
    \centering
    \includegraphics[width = 0.9\linewidth]{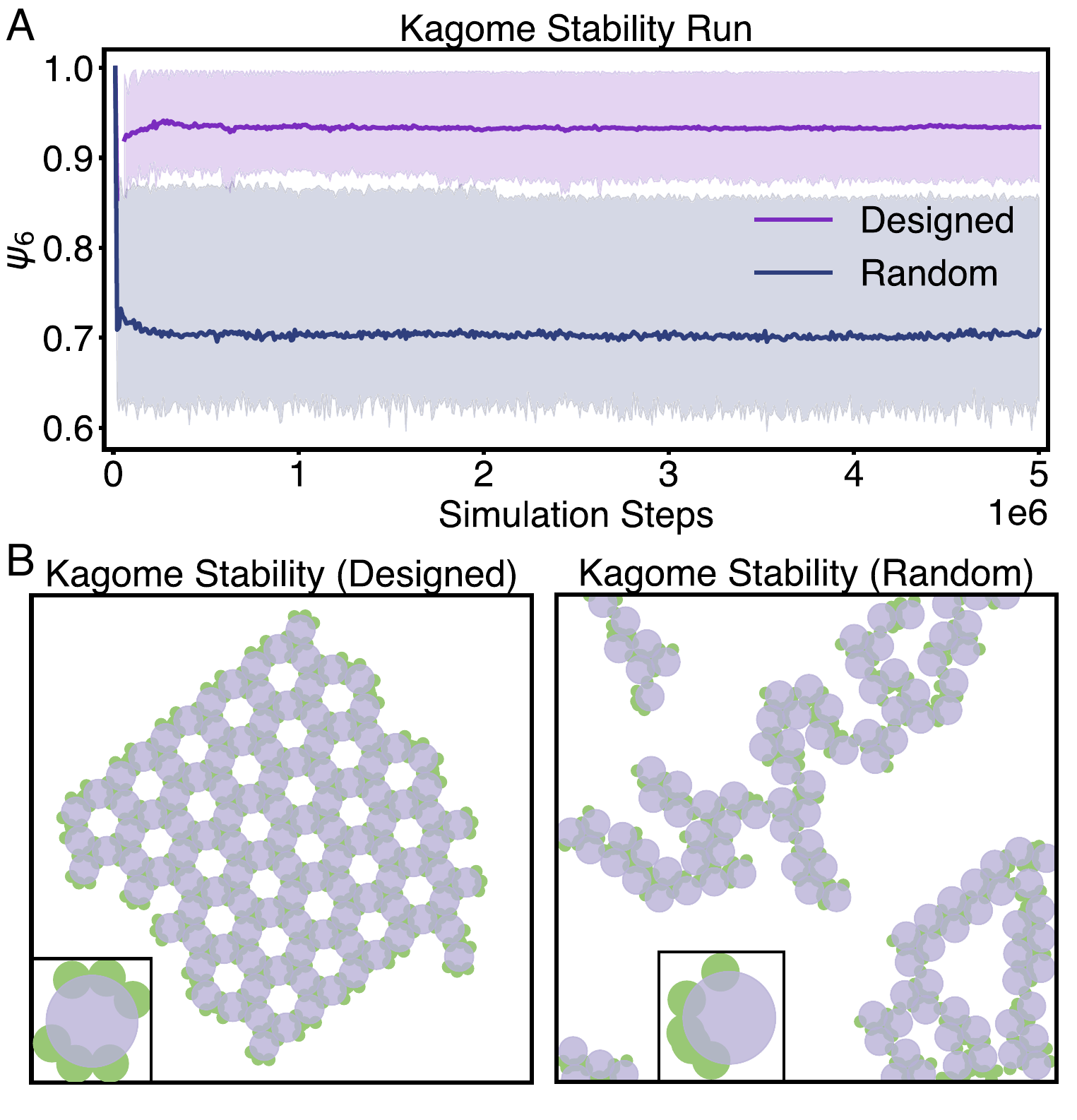}
    \caption{\textbf{Stabilizing a kagome lattice} A Stability of a system initialized in a kagome lattice configuration as a function of time. The purple (top) results use our optimized parameters, while the navy (bottom) results use random initial parameters. The optimization is initialized with random parameters. B Simulations run in HOOMD-blue demonstrating that the optimal parameters (left) stabilize the kagome lattice, while random parameters (right) cause the lattice to melt.}
    \label{fig:kagome}
\end{figure}

\section*{Results}
We demonstrate the optimization and design of patchy particles. These particles consist of a central particle plus a set of patches rigidly attached to the central particle (see Fig.~\ref{fig:model}A). The central particle describes the general shape of the patchy particle, and the set of patches governs the directional interaction of the patchy particle. Because we have enabled end-to-end differentiable simulations of anisotropic particles, we can directly optimize over the locations of the patches and the interaction matrices between patches. As a result, the rich design space available to patchy particles becomes feasible to search.

Here, we showcase three examples: (i) stabilization of a Kagome lattice, (ii) self-limiting 2D ring assembly and (iii) stabilization of 3D finite clusters. Designing parameters for any of these three examples using isotropic particles is highly challenging and requires complex pair potentials. By introducing anisotropy, we demonstrate robust design of each of these systems with simple interaction potentials.

In each example discussed, the patches interact via a Morse potential, and the central particles interact via a soft sphere potential or WCA potential.  

The three potentials are as follows:
\begin{itemize}
\item[] Morse Potential:
\begin{equation}\label{eq:morse}
    U(r) = \varepsilon (1 - e^{-\alpha (r - r_0)})^2
\end{equation}
\item[] Soft Sphere Potential:
\begin{equation}\label{eq:soft}
    U(r) = \varepsilon\left(\frac{\sigma}{r}\right)^\alpha
\end{equation}
\item[] WCA Potential:
\begin{equation}\label{eq:wca}
    U(r) = \varepsilon\left[\left(\frac{\sigma}{r}\right)^{12}-\left(\frac{\sigma}{r}\right)^6\right]+\varepsilon
\end{equation}
\end{itemize}

These interaction potentials are convenient for our use case, but can be readily changed for different applications. In each optimization example, we specify a loss function describing our targeted materials properties. This loss function must itself be fully differentiable, and thus cannot rely on discrete calculations.

\subsection*{Stabilizing a Kagome Lattice}

The Kagome lattice \cite{syozi1951statistics} is an open lattice structure (see Fig.~\ref{fig:model} and Fig.~\ref{fig:kagome}B) with a broad array of potential materials applications \cite{han2012fractionalized, xue2019acoustic}.  Self-assembling a Kagome lattice from isotropic potentials requires complicated potential landscapes with both attractive and repulsive wells \cite{pineros2016designing}, which are rarely possible to instantiate experimentally.  One way to simplify the assembly is to introduce anisotropy.  Inspired by the experimental realization of a Kagome lattice using Triblock Janus spheres \cite{chen2011directed}, we propose a general patchy particle model to optimize patch locations that stabilize a Kagome lattice, using a simple Morse potential. 

Each component in the model consists of a central particle and six patches that are rigidly attached to the central particle (Fig.~\ref{fig:kagome}B). The central particles interact with one another \emph{via} a soft sphere potential, and the patches each interact \emph{via} a Morse potential. We keep the Morse interactions fixed and optimize over the locations of the patches on the central particle. 

The optimization procedure follows the structure outlined in Method Section. We initialize the system in a Kagome lattice configuration with the orientations of the particles randomized. We then run 200 replicate \MD\ simulations. The simulations are run with for 40000 steps with a timestep ($dt$) of 1e-3 and 100 particles, at a temperature ($kT$) of 0.1 and an area fraction of 0.3 with periodic boundary conditions. We then measure the average loss function across the replicates. Because we initialize in the lattice configuration, the loss function for the optimization is simply the distance of the particles from their initial position: if the Kagome lattice is stable, the lattice will not melt and the positions of the particles will remain constant, up to vibrational motion. We compute the gradient of the average loss function with respect to the positions of the patches on the central particle. Lastly, we update the positions of the patches based on the value of the gradient using the Adam optimizer available in JAX. We repeat this procedure for 100 optimization steps with a learning rate of 0.1, and then starting from the optimal value, we run another 100 optimization steps with a learning rate of 0.05 and finally another 100 steps with a learning rate of 0.01. 

At the end of the JAX-based optimization procedure, we take the optimal parameters to run a longer simulation testing their stability using HOOMD-blue \cite{anderson2020hoomd, nguyen2011rigid, glaser2020pressure}.  Fig.~\ref{fig:kagome}A shows the $\psi_6$ order parameter measured using Freud \cite{freud2020} for designed building blocks and randomly generated ones.  We can see clearly that the designed building blocks stabilize the Kagome lattice and the parameters are transferable across different \MD\ engines.

\begin{figure}
    \centering
    \includegraphics[width = 0.9\linewidth]{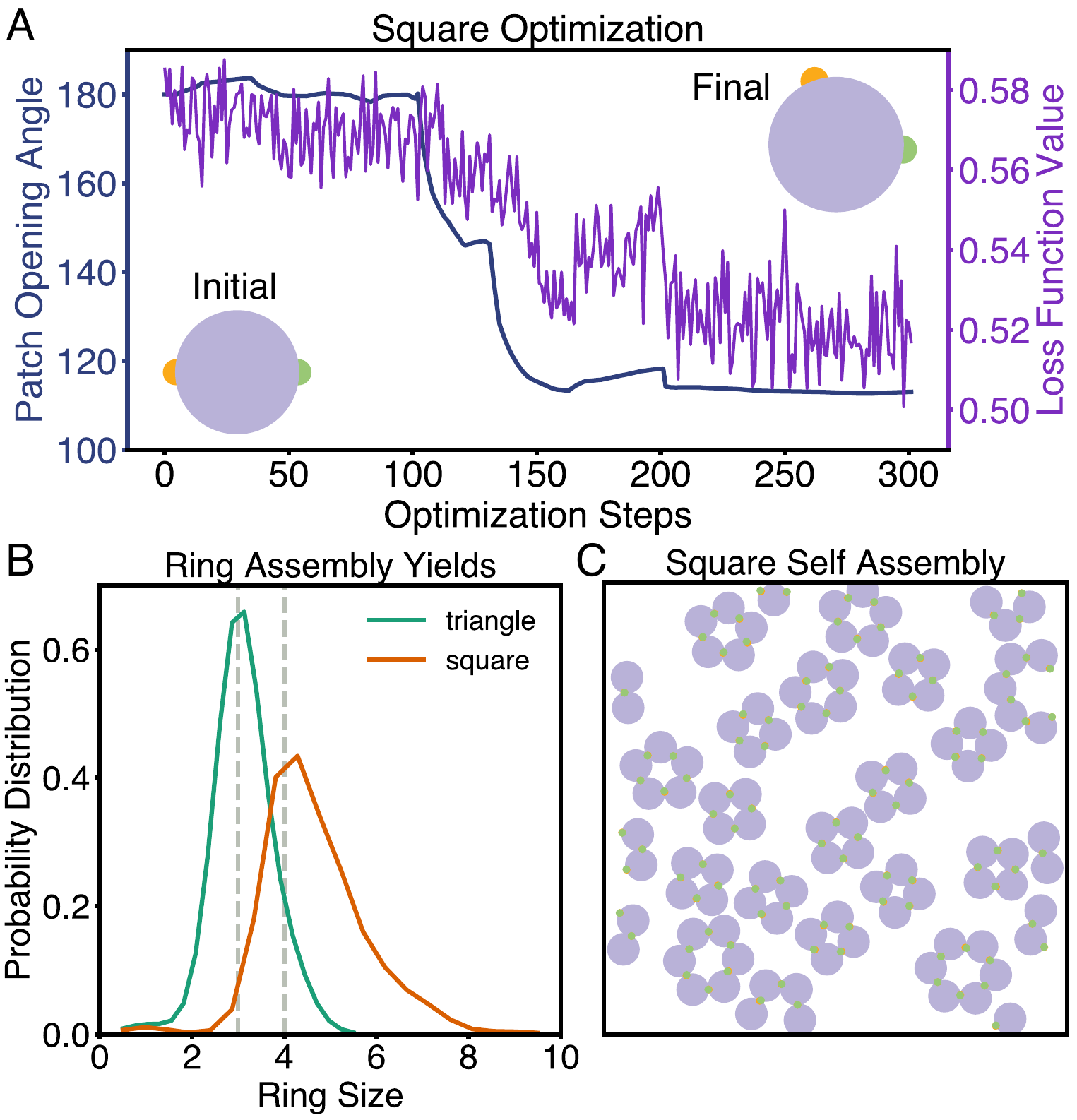}
    \caption{\textbf{Assembly of self-limiting rings.} A Optimization results for formation of square rings. The $x$-axis shows the optimization steps. The $y$-axis shows the patch opening angle on the left (navy), and the loss function on the right (purple). Over the course of the optimization, the patch opening angle tends towards a value slightly greater than 100 degrees. B Assembly yields computed from forward \MD\ simulations run in HOOMD-blue for the patchy particles designed to yield triangles (green) and squares (orange). The yield of each ring design peaks at the desired ring size, demonstrating that the design procedure was successful. C One example end result of assembling squares in a bath of particles. While some incorrect products (primarily larger rings) are observed, most of the particles are in square configurations.}
    \label{fig:ring}
\end{figure}

\subsection*{Self-Limiting Rings}
Despite the fact that natural systems rely on self-limited assembly, synthetically developing self-limiting structures remains a significant challenge \cite{hagan2021equilibrium}. Using our model, we design self-limited rings of varying sizes that self-assemble in a bath of components. 

To design self-limiting rings, we use a patchy particle model consists of a central particle and two patches. The central particles interact \emph{via} a soft sphere potential (Eq.~\ref{eq:soft}), and the patches interact \emph{via} a Morse potential (Eq.~\ref{eq:morse}). We optimize over both the location of the patches and the strength of the patch interactions. Each patch is allowed to vary independently. We initialize the patch positions and strengths randomly.

The optimization procedure again follows the structure listed in the Method section. Here, however, because we are interested in self-assembly rather than stabilization, we initialize the simulation with particles with random initial positions and orientations. 

To compute the loss for this calculation, we take the distance between each particle and its $M$ nearest neighbors ($M=3$ for square rings because squares consist of 4 particles, etc), and compare those distances to a reference structure. The reference structure is a perfectly assembled ring. We optimize over both the positions and the strengths of interactions of the patches. We compute the average loss over 128 replicate simulations that each run for 40,000 steps with a $d$t of 1e-3, at a temperature ($kT$) of 1.0 and an area fraction of 0.2. To reduce computational cost, we optimize over only the last 1,000 steps of the simulation. Despite this approximation, the gradients are meaningful enough to converge to optimal parameters.

Our model rapidly converges to a set of parameters that consistently forms independent rings of the specified size (Fig.~\ref{fig:ring}A). {\change Evolution of interactions between patches during optimization is included in the \SI.} While we do observe occasional malformed structures, we do not observe the formation of any extended structures in our system. We have thus successfully captured self-limiting behavior with differentiable patchy particles.

While one would naively assume that the optimal patch opening angle to form 4-component square rings would be 90 degrees, we find that the optimal opening angle is significantly wider, as shown in Fig.~\ref{fig:ring}. Our optimal results demonstrate a higher yield of square rings than the naive guess, as can be seen in Fig.~\ref{fig:ring}, as well as in the \SI. 

We hypothesized that the yield of squares is higher for a wider opening angle because it prevents the formation of triangles. To test this hypothesis, we performed two measurements. First, we measured the yield of triangles, squares, and pentagons in the system of particles designed to form squares. The results are given in Fig~\ref{fig:ring} and in the Supplementary Information section. We indeed observe that the formation of triangles is significantly suppressed for the designed parameters relative to the naive 90 degree guess. 

\subsection*{Self-Limiting 3D Clusters}

\begin{figure}[h]
    \centering
    \includegraphics[width = 0.9\linewidth]{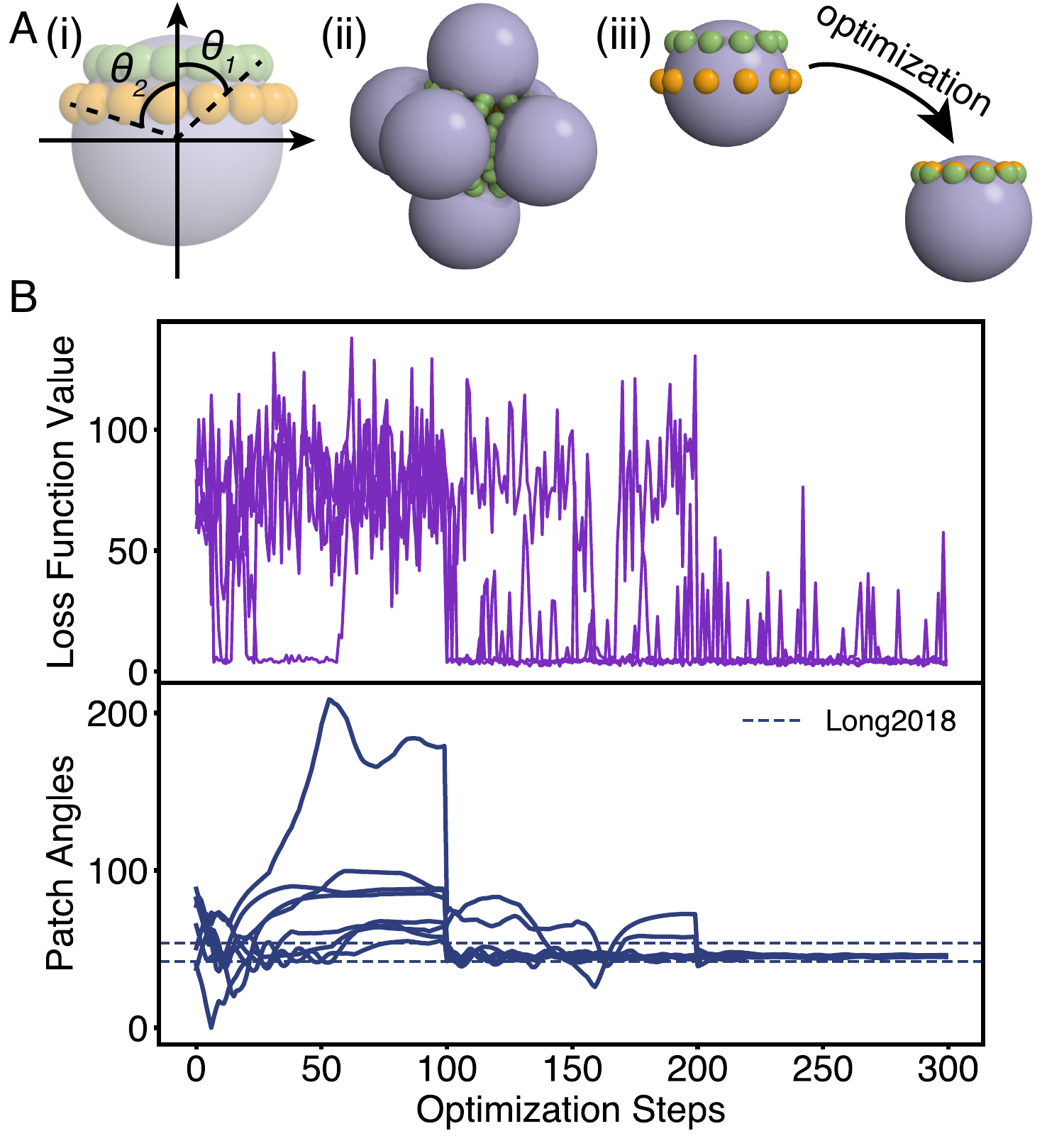}
    \caption{\textbf{Stabilization of octahedral cluster.} 
    (A) (i) patchy particle model used to optimize for octahedral cluster.  $\theta_1$ and $\theta_2$ determines the locations of the two rings of patches, where each ring has A-A type attractions. (ii) sample octahedral cluster (iii) patchy particle evolution from the beginning to the end of an optimization run.
    (B) Ensemble of optimization results for stabilization of an octahedral cluster. The $x$-axis shows the optimization steps. The $y$-axis shows the patch opening angle on the bottom plot (navy), and the loss function on the top plot (purple). The dotted lines show the optimized parameters from \cite{long2018rational}. Over the course of the optimization, the two patch angles converge to the same value, and fall into the range of the literature optimized values. The set of independent optimization runs all converge to the same optimal patch angles.
    }
    \label{fig:oct}
\end{figure}

While our 2D examples were successful, working with three-dimensional structures often poses different challenges. Inspired by virus shells, we demonstrate the stabilization of the simplest nontrivial platonic solid: the octahedron (Fig.~\ref{fig:oct}A(ii)). We leverage the non-isotropic interactions offered by patchy particles to find the patch positions and interaction strengths that consistently stabilize octahedral structures.

We begin with the model proposed by ~\cite{long2018rational}, consisting of two layers of patches in concentric circles (Fig.~\ref{fig:oct}A(i)). We optimize over the the positions of these circles of patches while fixing the interaction strength to be consistent with ~\cite{long2018rational}. Critically, though ~\cite{long2018rational} required mapping out regions of the free energy landscape to achieve assembly of clusters, we are able to recover features similar to their results with no explicit measurement of the free energies. 

Based on the simulation details in ~\cite{long2018rational}, we adapted the model to the \texttt{JAX-MD} simulation environment.  We use a WCA potential (Eq.~\ref{eq:wca}) for the center particle with $\sigma=5.0$ and $\varepsilon=1.0$ and a Morse potential for the patches with $\varepsilon=4.0$ and $r_0 = 0.0$. We simulate using a Langevin integrator with $\gamma = 5.0$, $d$t = 1e-4, $kT=0.8$, and a number density of 0.05.  Despite the modifications to the simulation parameters, the energy scale and dynamics of our system closely follow those described in ~\cite{long2018rational}.

Our optimization procedure for stabilizing 3D clusters closely mirrors our method in two dimensions, with minor modifications. We note that the self-assembly process for finite 3D clusters is considerably more complex from both thermodynamic and simulation perspectives. There are far more competing structures and the system takes much longer to equilibrate.  We mitigate these challenges by initializing our systems with 6 patchy particles, each consisting of 1 center particle and 20 patches, in a perfect octahedron. 

We again use a loss function that consists of computing nearest neighbor distances relative to those of a reference structure. In this case, the reference structure is the correctly formed octahedron. For every optimization step, we run 1 simulation for 200,000 steps and compute the gradient of the loss function to update patch locations.  The length of the simulation is determined by analyzing the time needed for to self-assemble a partial cluster (see \SI\ for details) and the number of replicates per optimization step is decided by gradient magnitude. We unexpectedly observed that using Langevin dynamics over Nosé-Hoover significantly reduces variation in the gradients. The whole optimization procedure consists of 300 optimization steps with three learning rates $[0.1, 0.05, 0.01]$ respectively using the Adam optimizer.

We initialize the optimization with randomly generated patch angle parameters (see Fig.~\ref{fig:oct}A(iii) as an example). With these random parameter values, the octahedron is not stable. This can be seen in the early values of the loss function: the loss at the outset of the optimization is both large and highly variable. As the optimization proceeds, the patch positions converge and the loss decreases. Ultimately, the optimization converges to parameters that reliably stabilize the three-dimensional cluster, as shown in Fig~\ref{fig:oct}B. We performed 4 independent optimizations for cluster stabilization, and for all the runs where the loss function converged, the patch locations converged to the same value also (see \SI\ for details).

Because we optimize for stabilization rather than assembly, our results deviate slightly from ~\cite{long2018rational}. In ~\cite{long2018rational}, the optimal parameters for octahedra assembly is $[42.0^\circ, 53.7^\circ]$, while our optimal parameters for octahedra stabilization is $[45.3^\circ, 46.0^\circ]$. We conclude that having two rings at similar locations is more favorable than having separated rings for the case of stabilization, and validate this conclusion with forward simulations of systems with each set of parameters. Indeed, our optimal parameters yielded a lower loss for stabilization than those in ~\cite{long2018rational} (see \SI\ for details). This may be alternatively explained by a difference in the two models: our patches have no volume of their own.

Despite these differences, our optimal ring positions fall between the two found in ~\cite{long2018rational}. The optimal location we find is closer to the inner ring in ~\cite{long2018rational}, which may indicate that the inner ring is a more vital feature in stabilization.  Additionally, we ran forward self-assembly simulations using the two sets of parameters using HOOMD-blue to test their ability to self-assemble, and the JAX-MD optimized parameters lead to a faster decrease of our loss function (see \SI\ for more details).

Critically, we were able to achieve these results without mapping the free energy landscape. This method not only provides a straightforward way to search the design space of anisotropic particles for properties of interest, but also showcase how small difference in model choice could lead to different optimal final results. This fast feedback loop for particle design could be instrumental in mapping to experimental systems.

\section*{Discussion}

We have introduced an end-to-end differentiable model system capable of capturing rich functional behavior in materials while still being simple enough to directly design. We have demonstrated the model by designing stabilization of an open lattice structure,  self-limiting assembly in 2D, and stabilization of 3D finite clusters.

In each case, we have made use of only one particle type. Previous efforts to, e.g., stabilize or assemble octahedral structures have relied on having $N$ different particle types to assemble a structure of $N$ particles \cite{zeravcic2014size, goodrich2021designing}. Though our individual components are more complex, the need to construct only one particle type renders our model system possible to manufacture at scale.

Though we believe our model offers significant potential for design of novel functional materials, its design potential is limited by {\change requiring differentiable loss functions and }computational expense. {\change To compute meaningful gradients based on the loss function, we can only use loss functions that do not rely on a sharp radial or nearest-neighbor cut-off. This limitation makes using traditional well-performed loss functions (order parameters), such as the local bond order parameter \cite{steinhardt1983bond}, less feasible and increases the difficulty of designing self-limiting structures, where one of the most straightforward loss functions is to count the number of particles in a cluster. The limitation on loss functions is both a challenge and an opportunity.  With more machine learning techniques being incorporated in materials design, we need to rethink how we describe materials structures and properties and come up with new robust and accurate descriptors that fit the inverse-design method at hand.}

{\change On the side of computational expense, our method is limited mostly by GPU memory.} Large systems of particles yield highly memory-intensive gradient computations, so smaller systems or behaviors that are well-described by local structure are better-suited to the model. {\change The memory usage for a particular optimization run is decided by the combination of system size ($N$), run timesteps ($r_s$), and batch size ($b$). For a GPU with 32Gb of memory, we can run a simulation of $N\times r_s\times b\leq 10^8$. The limitation can be mitigated in a few ways: run on a GPU with higher memory, distribute batch sizes on multiple GPUs, and distribute a single big system on multiple GPUs.  Currently, there are GPUs that have higher memory capacity, and JAX-MD already allows distributing batch sizes on multiple GPUs. We are actively working on parallelizing a single system on multiple GPUs to reduce the memory limitation.}

We have only begun to explore the range of behaviors available to this model. One possible extension of our work on self-limited structures is to learn the rules of assembly for larger, multi-component Virus-Like Particles (VLPs), biomimetic shells that have the potential to be used for drug delivery. Additionally, while we made use of zero width patches, if we instead considered patches that were physical particles, we could explore the realm of colloidal molecules. These structures have been instantiated experimentally, but the design space of colloidal molecules is vast and under explored \cite{glotzer2007anisotropy, walther2013janus, hueckel2021total}.

This model is the right paradigm to reach the longstanding goal of directly designing for functional behavior by optimizing small, simple components. 

\acknow{We thank Carl Goodrich, Ryan Krueger, and Megan Engel for helpful discussions. This material is based upon work supported by the Office of Naval Research (ONR N00014-17-1-3029), the National Science Foundation Graduate Research Fellowship under Grant No. DGE1745303, NSF Grant DMR-1921619, and the NSF AI Institute of Dynamic Systems (\#2112085).  3D particles and trajectories are rendered by INJAVIS \cite{engel_michael_2021_4639570}. Data generated in this paper is managed by Signac \cite{signac_commat, signac_scipy_2018}. }

\showacknow{} 

\bibliography{references}

\end{document}